\documentclass[hidelinks]{article}

\usepackage{arxiv}

\usepackage{doi}

\usepackage{graphicx}
\usepackage{enumerate}
\usepackage{natbib}
\usepackage{url}
\usepackage[misc]{ifsym}
\usepackage{amsmath}
\usepackage{multirow}
\usepackage{cleveref}       
\usepackage{calc}

\title{Fairness Score and Process Standardization: Framework for Fairness Certification in Artificial Intelligence Systems}

\author{ \href{https://orcid.org/0000-0003-4553-5861}{\includegraphics[scale=0.06]{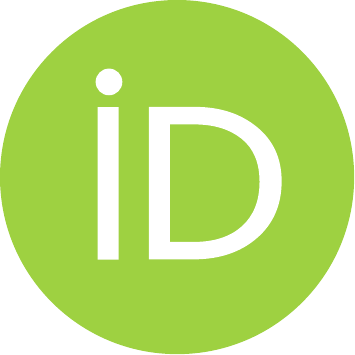}\hspace{1mm}Avinash ~Agarwal} \\
	Telecommunication Engineering Centre\\
  Ministry of Communications\\
  New Delhi, India \\
  \texttt{avinash.70@gov.in} \\
	\And
	\href{https://orcid.org/0000-0003-3193-1521}{\includegraphics[scale=0.06]{orcid.pdf}\hspace{1mm}Harsh ~Agarwal} \\
	Adobe Inc.\\
  Noida, India \\
	\And
	\hspace{1mm}Nihaarika ~Agarwal \\
	Manipal Institute of Technology \\
  Manipal, India \\
}

\renewcommand{\undertitle}{}
\renewcommand{\headeright}{}
\renewcommand{\shorttitle}{Fairness Score and Certification Standardization in AI Systems}

\hypersetup{
pdftitle={Fairness Score and Process Standardization: Framework for Fairness Certification in Artificial Intelligence Systems},
pdfsubject={q-cs.AI},
pdfauthor={Avinash ~Agarwal, Harsh ~Agarwal, Nihaarika ~Agarwal},
pdfkeywords={Bias, Fairness Score, Standardization, Artificial Intelligence, Ethics, Audit, SOP},
}

\begin{document}
\maketitle

\begin{abstract}
Decisions made by various Artificial Intelligence (AI) systems greatly influence our day-to-day lives. With the increasing use of AI systems, it becomes crucial to know that they are fair, identify the underlying biases in their decision-making, and create a standardized framework to ascertain their fairness. In this paper, we propose a novel Fairness Score to measure the fairness of a data-driven AI system and a Standard Operating Procedure (SOP) for issuing Fairness Certification for such systems. Fairness Score and audit process standardization will ensure quality, reduce ambiguity, enable comparison and improve the trustworthiness of the AI systems. It will also provide a framework to operationalise the concept of fairness and facilitate the commercial deployment of such systems. Furthermore, a Fairness Certificate issued by a designated third-party auditing agency following the standardized process would boost the conviction of the organizations in the AI systems that they intend to deploy. The Bias Index proposed in this paper also reveals comparative bias amongst the various protected attributes within the dataset. To substantiate the proposed framework, we iteratively train a model on biased and unbiased data using multiple datasets and check that the Fairness Score and the proposed process correctly identify the biases and judge the fairness.
\end{abstract}


\section{Introduction}
\label{intro}
Artificial Intelligence (AI) services are used widely in various fields and aspects such as finance, education, medicine, and law. Predictive systems use large datasets for training and several algorithms to enable accurate and fast decision-making. These systems make decisions based on their learnings from the training data and algorithms used. The systems are improving continuously, becoming more complex and automated. As such, manual supervision and validation of their outcomes are reducing. Thus it is pertinent to ask some fundamental questions, such as 1) how the system reaches the outcome? 2) how often the system's outcomes are wrong 3) is the system being unfair to a particular group? \citep{sharma2019certifai}.

Over the centuries, discrimination based on race, gender, religion, etc., has existed in society and human psychology. The assumption that the decisions made using AI systems will be fair may not be correct. These systems learn from datasets consisting of real-world events and observations. The biases or prejudices present in the real world seep into the systems through various steps in model building. Hence this discrimination can be observed in the results and decisions made by the AI systems. The discrimination and discriminatory decisions can be termed as biases and biased decisions, respectively. Biased decisions hinder the rights and sentiments of the unprivileged group and may cause a negative social impact. Various search engines and recommender systems had race and gender discrimination \citep{leavy2020data}.

Thus it is necessary to detect biases in AI systems and then mitigate them so that these systems are fair. However, how do the users know whether a decision involving them is made using a fair system or a biased system? Can the developing organization's claim to fairness be taken at face value? Are there metrics/ methodologies to verify the claim? Can some audit processes be standardized such that neutral certifying agencies can audit and certify AI systems for fairness? We answer some of these questions in this paper. The aim of this study is to assess the fairness of a given AI system and to enable third-party audit and certification for fairness of AI systems. The scope does not however include recommending any corrective actions to be taken if the AI system under analysis does not meet the qualifying criteria.

\paragraph{Our contributions:}
\begin{enumerate}
    \item We propose a Fairness Score to measure the overall fairness of an AI system. We also propose Bias Index to measure the bias of each protected attribute in a dataset.
    \item We propose a Standard Operating Procedure (SOP) for issuing Fairness Certification for data-driven Artificial Intelligence systems that would provide a framework to operationalize the concept of fairness and facilitate commercial deployment of such systems.
    \item We explain the need for third party Fairness Certification for such systems.
    \item We define the desired characteristics of metrics to detect bias at different phases of such AI systems.
    \item We propose the metrics out of a plethora of such metrics sufficient to detect bias.
\end{enumerate}

\paragraph{Methodology:}
We review the various cases of bias in AI systems and the different ways to define fairness. Then, we shortlist the metrics that properly detect bias in a training dataset and a trained model. Accordingly, we propose the Fairness Score, Bias Index, and an SOP for detecting bias. Finally, we establish our proposals by applying them to a model, which is trained multiple times using the following popular open datasets one by one:
\begin{enumerate}
  \item UCI Adult dataset (four scenarios):
  \begin{enumerate}
    \item Original dataset
    \item Dataset after balancing for gender
    \item Dataset after balancing for race
    \item Dataset after balancing for marital status
  \end{enumerate}
  \item German Credit dataset (two scenarios):
  \begin{enumerate}
    \item Original dataset
    \item Dataset after balancing for gender
  \end{enumerate}
  \item Health Insurance Cross-Sell Prediction dataset (one scenario)
  \begin{enumerate}
    \item Original dataset
  \end{enumerate}
\end{enumerate}

\paragraph{Related work:}
Various researchers have worked in areas related to biases in AI systems, concepts of fairness, and metrics to measure fairness, as elaborated in subsequent sections. Most of these works are related to theoretical or mathematical aspects. Our proposal is different as we focus on the practical implementation aspects aiming to standardize the audit procedure to check the fairness of an AI system to enhance its trustworthiness. While most researchers have proposed different metrics to check for biases, we consider these metrics and present an integrated Fairness Score. Our work would be found beneficial by organizations developing AI systems as well as those deploying such systems. It would also facilitate benchmarking and independent third party audit for fairness.

\section{Need for Fairness Certification}
\label{needForCertification}
Bias creates unwanted drifts between different groups. It gives an advantage to the people belonging to one particular class irrespective of other vital parameters that value that decision-making process. Apart from this, it also affects the business and its profits based on the predictive results of the system. Therefore, ensuring the trustworthiness of AI systems is not only a research subject but also a business requirement \citep{curionin2019slides}. Moreover, AI bias could negatively affect every sector as AI systems become widespread \citep{aifairness}.

Previous studies have highlighted cases where biases in AI-driven systems have caused unwanted impacts and results. For example, Amazon developed an online AI-based recruitment system where the candidates were shortlisted based on their resumes. By 2015, it came to the company’s attention that ratings were biased in favour of male applicants for technical job titles like software developers and architects. On investigation, it was found that the data used for training the system was having gender bias as it consisted of resumes of mostly male employees in reflection to the then actual hiring trend in the company  \citep{kodiyan2019overview}. Research on online freelance marketplace based on evidence from TaskRabbit and Fiverr revealed that social feedback on these sites often has a significant statistical relationship with perceived gender and race \citep{hannak2017bias}. Various facial recognition systems have biases and give accurate detection results only for people of a certain race and often incorrect results for the other demographic groups \citep{garvie2016facial}. The US healthcare system was observed to be racially biased and preferred treating white patients over the ones with African origins \citep{obermeyer2019dissecting}. The Correctional Offender Management Profiling for Alternative Sanctions (COMPAS) score, used in States such as California and Florida, was reported to be biased against African origin inmates according to certain measures of fairness \citep{wadsworth2018achieving}. Princeton research showed that off-the-shelf AI software was biased in associating words like `woman' and `girl' with arts as compared to science or math \citep{hadhazy2017biased}. Studies showed that online search queries were more likely to return ads from services related to arrest records when names of a particular race were searched than when names of the other race were searched \citep{sweeney2013discrimination, lee2019algorithmic}. Biases in expert systems can also lead to the false detection of disease, wrong prescription of medical tests and more visits to the hospital for routine check-ups \citep{zheng2017resolving}.

These and many similar kinds of research indicate the negative impact of biases in AI systems and the need for ensuring fairness in such systems. As these systems get embedded in day-to-day applications and essential services such as healthcare, judicial, and many more, it becomes pertinent for users to know that the system they are using is fair. Small service providers to large corporations and government organizations use AI/ Machine Learning (ML) based applications to provide services. However, everyone does not have the resources or technical know-how to test the fairness of the AI systems they are deploying. The AI application could be an off-the-shelf product or a customized product developed specifically for that end-user. There is, therefore, a need for a standard testing procedure to test the AI application for biases. A Fairness Certificate issued by a neutral certifying authority based on such standard testing procedure would assure the users of the fairness of the AI/ ML system.

The standard testing procedure should protect proprietary AI system developers, privacy, and security of the training dataset \citep{segal2021fairness}. It should comply with the existing national and international regulations \citep{tommasi2021towards}. Finally, it should not only protect the rights of individuals but also help the competition to grow and provide the best services \citep{schubert2019economy}.

\section{Causes of Bias in AI Systems}
\label{causesOfBias}
Biases can enter an AI system at any point in its cycle. Empirical findings have shown that data-driven methods can unintentionally encode existing human biases and introduce new ones \citep{chouldechova2018frontiers}. According to \citep{ferrer2021bias}, three causes of biases are:

\paragraph{Bias in modelling:} These biases correspond to the algorithmic biases, caused due to the algorithms defined for model functioning and decisive actions. Algorithms often reflect biases of the developers and hence the society at large \citep{shah2018algorithmic}. Also, they do it at a potentially massive scale and without due oversight. \citep{panch2019artificial}.

\paragraph{Bias in training:} The systems learn based on the data used to train them. The biases existing in society are often manifested in the data used by AI algorithms and can be modelled and formally defined \citep{ntoutsi2020bias}. The data may represent the ideologies of only one group of the population or specific cases, which introduces prejudices or biases in the decisions.

\paragraph{Bias in usage:} These biases arise when the models or systems are used for a purpose not defined for them. The AI systems designed for making decisions for a specific group or section provide biased results when used for any other group. Misinterpretation of the outcomes of the system also causes wrong and indeliberate actions or biases. Biases also creep in when the systems are used for intended purpose but one specific class has more use instances while the AI system keeps learning from its usage.

\section{How to define Fairness}
\label{fairnessDefinition}
While we create a certification framework and identify fairness metrics, it is necessary to define fairness. AI fairness is a rapidly growing topic of inquiry \citep{hajian2016algorithmic}. Multiple definitions of fairness have been proposed \citep{verma2018fairness, zliobaite2017fairness, mehrabi2021survey}. Some of these definitions were proposed for formalizing fairness from other disciplines \citep{hutchinson201950}. In some cases defining fairness comes as a legal compulsion, for example, lending money through The Equal Credit Opportunity Act in the United States \citep{arnold2019factsheets}.

Fairness can be broadly defined as the state of being impartial towards every individual and group involved. However, fairness can be perceived differently by different people and contexts \citep{mulligan2019thing}. Thus it is not easy to have a single definition of fairness for all AI systems.

Various researchers have come up with their definitions for fairness in AI systems. The outcomes would differ depending on the difference in the aspects of fairness focused on by different definitions of fairness \citep{aifairness}.

For applying fairness well, one should understand it within a specific context. Guiding questions to determine fairness could be: Are there particular groups that may be advantaged or disadvantaged by the algorithm/ system in that context? In case of uncertainty or errors in the AI system, what is the potential adverse effect to different groups? \citep{fatml}

Individual fairness implies that if two individuals differ only in protected attributes such as gender or race, they should receive similar outcomes. There is individual discrimination if they receive different outcomes \citep{aggarwal2019black}. A distance metric may be used to measure the similarity between the two individuals \citep{dwork2012fairness}.

Group fairness involves dividing the data into different categories based on the protected attributes and ensuring that the outcomes are similar for all the groups. Group fairness averages out the data of individuals in a group, so it may or may not represent the correct picture.

Other than individual and group fairness, fairness definitions can also be classified using the statistical approach used: definitions based on the predicted outcome; definitions based on predicted and actual outcomes; and definitions based on predicted probabilities and actual outcomes \citep{verma2018fairness}. 

Similarity-based definitions use non-sensitive attributes also, unlike the above fairness definitions.

Counterfactual fairness uses plotting a graph to check whether the outcome-defining attributes are correlated to the sensitive attributes \citep{kusner2017counterfactual, ntoutsi2020bias}.

These definitions are independent of each other, and multiple definitions can be applied to an AI system simultaneously. It is also important to note that for an AI system/ dataset to be fair, it does not mean that it should always have equal representation for all the classes across the different demographics. Equality should be in sync with the ground reality.

While many formal mathematical definitions of fairness have been proposed, formalizing fairness is still a problem. These definitions have statistical limitations, and there are adverse effects of enforcing such fairness measures on group well-being \citep{corbett2018measure}. Thus, there might be systems that do not meet some of the definitions but still are fair. It is therefore essential to understand the context of the AI system before judging it for fairness. Obviously, fairness is a multi-disciplinary topic. \citep{romei2014multidisciplinary} provides a survey of fairness analysis in multiple disciplines. 

In the subsequent sections, we define the features of a good fairness metric and discuss some of the fairness metrics along with their mathematical representations.

\section{Desired Characteristics of Fairness Metrics}
\label{desiredCharacteristicsFairnessMetrics}
We need to measure a data-driven AI system against some fairness criteria or metrics to evaluate whether the system is fair or not \citep{hertweck2021moral}. Unwanted biases in training data, due to either prejudice in labels or under-/ over-sampling, can be checked using statistical tests on datasets or models \citep{barocas2016big, bellamy2018ai}.

While the fairness definitions and computation of fairness metrics are explicit, applying inflexible thresholds on such metrics is not practical as the interpretation of the numbers is often subjective. As such, the thresholds should be decided considering the constraints in view. Before computing these metrics, it is crucial to define the characteristics of a good fairness metric.

A good fairness metric should be clear in terms of its mathematical and theoretical significance and also in terms of features, outcomes, and situations it takes into consideration during evaluation. There should be clear guidelines on the metric review to draw valuable inferences about the biases in the system. Its constraints, drawbacks, and advantages should be known to enable its selection according to the application of the system. 

A good fairness metric is expected to be quantitatively and accurately measurable to calculate the bias present in the AI system. Furthermore, it should assist in coming up with effective mitigation processes for the biases and setting up fairness benchmarking guidelines for an AI system.

An excellent metric should be universally accepted for the systems used in various parts of the world. This is required to avoid discrepancies in the fairness definitions or guidelines for a system in various geographies and to establish uniform justice among people around the globe.

As mentioned in previous section, the context of fairness is domain dependent and there maybe inherent trade-off in aggregating multiple metrics in different use cases \citep{kleinberg2016inherent, singh2021fairness}.

\section{Standard Fairness Metrics}
\label{standardFairnessMetrics}
Here we have explained some of the standard fairness metrics which abide by the above characteristics. The various notations used in their mathematical representations are:
\begin{align*}
Y &: training \ label \\
Y' &: predicted \ label \\
S \in \{0, 1\} &:  binary \ indicator \ of \ the \ protected \ class \\
&\ \ 1 \ indicates \ the \ privileged \ group \\
&\ \ 0 \ indicates \ an \ unprivileged \ group
\end{align*}

\subsection{Demographic Parity}
\label{demographicParity}
It is a measure of whether the probabilities of favourable outcomes for privileged and unprivileged groups of data in a protected attribute are the same.

It is independent of the prediction/ system outcome, making it simpler and easier to implement. However, a drawback to this metric is that it does not consider the genuine real-world differences between the privileged and unprivileged groups. Thus, this metric will be inappropriate if the reason for the difference in probabilities of favourable outcomes for different groups in a protected attribute is legitimate, for example, distribution of breast cancer probability based on gender \citep{hinnefeld2018evaluating}.
For an unbiased system, it can be mathematically represented as:
\begin{equation}
P(Y = 1 | S = 0) = P(Y = 1 | S = 1)
\end{equation}
In case of bias, the two sides of the equation will not be equal. This metric can be calculated in two different ways:

\subsubsection{Statistical Parity Difference}
\label{statisticalParityDifference}
It is the difference of the probabilities of favourable outcomes for unprivileged and privileged classes in a protected attribute and represented as:
\begin{equation}
P(Y = 1 | S = 0) - P(Y = 1 | S = 1)
\end{equation}
For an unbiased system, its value should be zero.

\subsubsection{Disparate Impact}
\label{disparateImpact}
It is the ratio of the probabilities of favorable outcomes for unprivileged and privileged classes in a protected attribute \citep{lohia2019bias} and represented as:
\begin{equation}
\frac{P(Y = 1 \| S = 0)}{P(Y = 1 \| S = 1)}
\end{equation}
For an unbiased system, its value should be one.

\subsection{Equal Opportunity}
\label{equalOpportunity}
It measures the equality of the true positive rates, i.e. equality of recall value for the privileged and unprivileged groups. It simply means that every group should have equal opportunities to be successful. It is also called positive rate parity.
Mathematically, it can be written as \citep{hardt2016equality}:
\begin{equation}
P(Y' = 1 | S = 1, Y = 1) - P(Y' = 1 | S = 0, Y = 1)
\end{equation}
Its disadvantage is that it requires large labelled historical data \citep{awwad2020exploring}.

\subsection{Equal Mis-Opportunity}
\label{equalMisOpportunity}
According to this metric, the accuracy measured as false positive rate should be the same for privileged and unprivileged groups. It is also known as predictive equality \citep{corbett2017algorithmic}. 
Mathematically, it can be written as \citep{hinnefeld2018evaluating}:
\begin{equation}
P(Y' = 1 | S = 1, Y = 0) - P(Y' = 1 | S = 0, Y = 0)
\end{equation}

\subsection{Average Odds}
\label{averageOdds}
It is the combination of equal opportunity and equal mis-opportunity metrics. It measures whether the true positive and the false positive rates are the same for privileged and unprivileged groups. It simply means that the favourable outcome is independent of the protected feature.
Mathematically, it can be written as:
\begin{equation}
(P(Y' = 1 | S = 0, Y = y) - P(Y' = 1 | S = 1, Y = y))/2
\end{equation}
where,
\begin{align*}
y \in \{0, 1\}
\end{align*}
It tries to ensure high accuracy in all demographics \citep{hardt2016equality}. It can be preferred when discrimination can be accepted as long as it is justified by actual trustable data \citep{castelnovo2021zoo}.

\subsection{Distance Metrics}
\label{distanceMetrics}
These metrics are used to calculate the distance between two points. In terms of fairness metrics, these metrics can be used to evaluate the similarity between two data points. These can also be used to compare two datasets, for example, for measuring the change between an original dataset and the dataset after distortion/ bias mitigation. 

Some familiar distance metrics are Euclidean distance, Manhattan distance, and the more generalized Minkowski distance.

One disadvantage of these metrics is that large valued variables might influence results. Also, they are not suitable for image classification or document classification \citep{pandit2011comparative}. Finally, these metrics are also not suitable to measure similarities between groups within the same dataset.

Some researchers have suggested using distance metrics to measure bias while focusing on using these metrics to determine individual fairness. In our experiment, we consider group fairness and hence have not used the distance metrics to calculate bias. However, the fairness certification framework proposed in this paper can be extended for individual fairness also.

\section{Proposed Framework for Fairness Certification}
\label{framework}

\subsection{Fairness Score and Bias Index}
\label{fairnessScoreAndBiasIndex}
Different users might use different metrics to check the fairness of an AI system. Hence, it is crucial to standardise the bias measurement on a linear scale so that a uniform scale can be used to assess fairness and enable the comparison of different AI systems. Therefore, we introduce Bias Index for each protected attribute and Fairness Score for the overall system as the standard benchmarks for measuring fairness.

Bias Index is defined for each protected attribute in the system as follows:
\begin{equation}
BI\textsubscript{i} = \sqrt{\frac{\sum_{j=1}^{n} (M\textsubscript{ij} - M\textsubscript{j}')\textsuperscript{2}}{n}}
\end{equation}
where,
\begin{align*}
i &: number \ of \ the \ protected \ attribute \\
j &: number \ of \ the \ fairness \ metric \\
n &: total \ number \ of \ fairness \ metrics \ used \\
m &: total \ number \ of \ protected \ attributes \ considered \ in \ the \ AI \ system \\
M\textsubscript{ij} &: value \ of \ the \ j\textsuperscript{th} \ fairness \ metric \ for \ the \ i\textsuperscript{th} \ protected \ attribute \\
M\textsubscript{j}' &: ideal \ value \ of \ the \ j\textsuperscript{th} \ fairness \ metric \\
&\ \ i.e. \ 0 \ for \ difference \ metrics \ and \ 1 \ for \ ratio \ metrics
\end{align*}
Fairness Score is defined for the AI system as follows:
\begin{equation}
FS = 1 - \sqrt{\frac{\sum_{i=1}^{m} (BI\textsubscript{i})\textsuperscript{2}}{m}}
\end{equation}
Substituting the equation for BI, we get,
\begin{equation}
FS = 1 - \sqrt{\frac{\sum_{i=1}^{m} \sum_{j=1}^{n} (M\textsubscript{ij} - M\textsubscript{j}')\textsuperscript{2}}{mn}}
\end{equation}
While Bias Index corresponds to the degree of bias for a particular protected attribute in a dataset or model, Fairness Score corresponds to the degree of fairness in the entire model taking into account all the protected attributes together. For a fair system, Bias Index for each protected attribute should be zero, whereas the Fairness Score should be one. AI systems may have more than one protected attribute, and so there would be as many Bias Indexes as the number of protected attributes, but there will be only one Fairness Score for the model. The AI system might be biased for some of the protected attributes and fair for the other protected attributes, which the corresponding Bias Indexes would reflect; the Fairness Score, on the other hand, would reflect the overall fairness.

\subsection{Proposed Standard Operating Procedure for Fairness Certification}
\label{fairnessCertificationSOP}
Considering the desired characteristics of the fairness metrics and their applicability
at different stages of model building, we propose a Standard Operating Procedure for issuing Fairness Certificate for data-driven AI Systems as follows:
\begin{enumerate}
    \item Identify the protected attributes in the dataset used in the AI system.
    \item Identify the privileged and unprivileged classes for each attribute in the dataset.
    \item Define the tolerance band for each attribute. Ideally, it should remain fixed for proper benchmarking.
    \item Check the training dataset for fairness. Compute values of Statistical Parity Difference and Disparate Impact (or any other metric considered appropriate) for each protected attribute of the training dataset. Check if the values fall in the tolerance band.
    \item Train the AI system using the training dataset.
    \item Run the system on a test dataset and tally the outcomes.
    \item Compute values of Equal Opportunity Difference, Equal Mis-Opportunity Difference, and Average Odds Difference (or any other metric considered appropriate) for each protected attribute of the outcome. Check if the values fall in the tolerance band.
    \item Plot the various metrics on a graph and see if all the metrics are within the tolerance band.
    \item Calculate the Bias Index for each protected attribute and the Fairness Score for the overall AI system.
    \item Based on the above observations, the fairness certification can be issued.
    \item In a real-world scenario, the developers might not share the training dataset with the certifying agency. In such cases, steps 4 and 5 have to be skipped. So, the test dataset should be close to unbiased for all the protected attributes. Any deviation of the metrics computed on the outcomes from the mean position would indicate bias in the system. In such cases, the certificate issued by the certifying agency should mention that the statistical parity and disparate impact have not been checked due to non-availability of the training dataset. Further, a legal framework could be considered where, depending on the requirements and use case, the developers can be mandated to share their training datasets with the certifying agency for certification.
    \item As the system is operationalized and keeps learning from real-world data, it is necessary to check periodically for any biases introduced after the initial tests. As such, periodic recertification is recommended.
\end{enumerate}

\section{Experiment}
\label{experiment}
We evaluated our proposed framework on seven different datasets related to the popular open datasets: UCI Adult dataset (four scenarios), German Credit dataset (two scenarios), and Health Insurance dataset. For each dataset, we performed the following steps:

We identified the protected attributes and also the privileged and unprivileged classes for each such attribute. Then, we cleaned the dataset by removing special characters, dropping some skewed attributes, and transforming the categorical attributes using one-hot encoding.
We randomly divided the dataset into 85\% training set and 15\% test set. We used RandomForestClassifier to train the model and make predictions. We calculated the following fairness metrics for each protected class: 
Statistical Parity Difference (SPD) and Disparate Impact (DI) for training dataset;
Equal Opportunity Difference (EOD), Equal Mis-Opportunity Difference (EMOD), and Average Odds Difference (AOD) for test set predictions using the trained model.

We used visualization tools to plot these observations. We finally calculated the Bias Index for each protected attribute and the Fairness Score for the model.
We repeated these for all seven cases as mentioned above.

\subsection{UCI Adult Dataset}
\label{UCIAdultDataset}
UCI Adult dataset is also known as the Census Income dataset. The data has been extracted from the 1994 Census dataset. It is used to predict whether an individual's income exceeds \$50k per annum or not. The dataset contains attributes like age, gender, marital status, occupation, among others.

As part of pre-processing, we reclassified the classes in the marital-status attribute as following:
\begin{align*}
Married &= \{'Married-civ-spouse', \ 'Married-AF-spouse'\} \\
Unmarried &= \{'Widowed', \ 'Never-married', \ 'Divorced', \\ &\ \ 'Separated', \ 'Married-spouse-absent'\}
\end{align*}

For our experiment, we considered the following configurations:
\begin{align*}
Protected \ attributes&: gender, \ race, \ marital-status \\
Privileged \ class&: Male \ (gender), \ White \ (race), \\ &\ \ Married \ (marital-status) \\
Unprivileged \ class&: Female \ (gender), \ Other \ than \ White \ (race), \\ &\ \ Unmarried \ (marital-status) \\
Favourable \ outcome&: Income > \$50k/yr
\end{align*}

Marital status has been included as a protected field because on analysis of the UCI Adult Dataset, a clear bias was observed towards married individuals as compared to individuals who were not in a relationship. One possible reason could be the belief at that time that married individuals were considered more stable and safer bet. The first iteration of model training and bias calculation was done on the original (biased) dataset. For the next iteration, we used oversampling (SMOTENC) and random undersampling to mitigate bias in the race attribute. We balanced the number of favourable outcomes for both the privileged and unprivileged classes. A fresh model was trained on this resampled dataset, and a new set of fairness metrics was calculated for each protected attribute. We repeated the same steps two more times, balancing gender one time and marital status the other.

\begin{figure}[h]
\centering
\includegraphics[width=\linewidth]{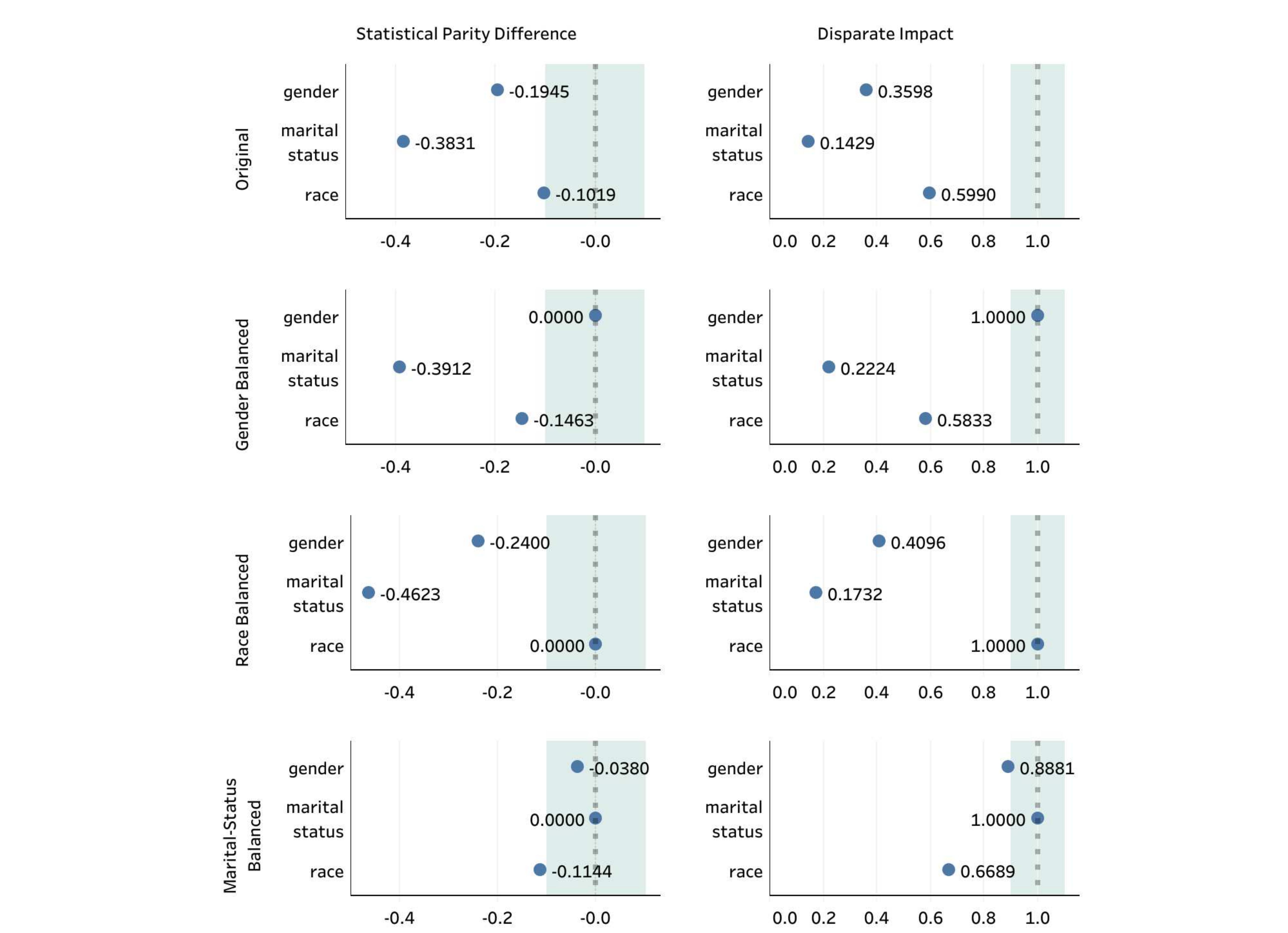}
\caption{Statistical Parity Difference and Disparate Impact values for the pre-training datasets for the four scenarios of the UCI Adult dataset}
\label{fig:1}
\end{figure}

\begin{figure}[h]
\centering
\includegraphics[width=\linewidth]{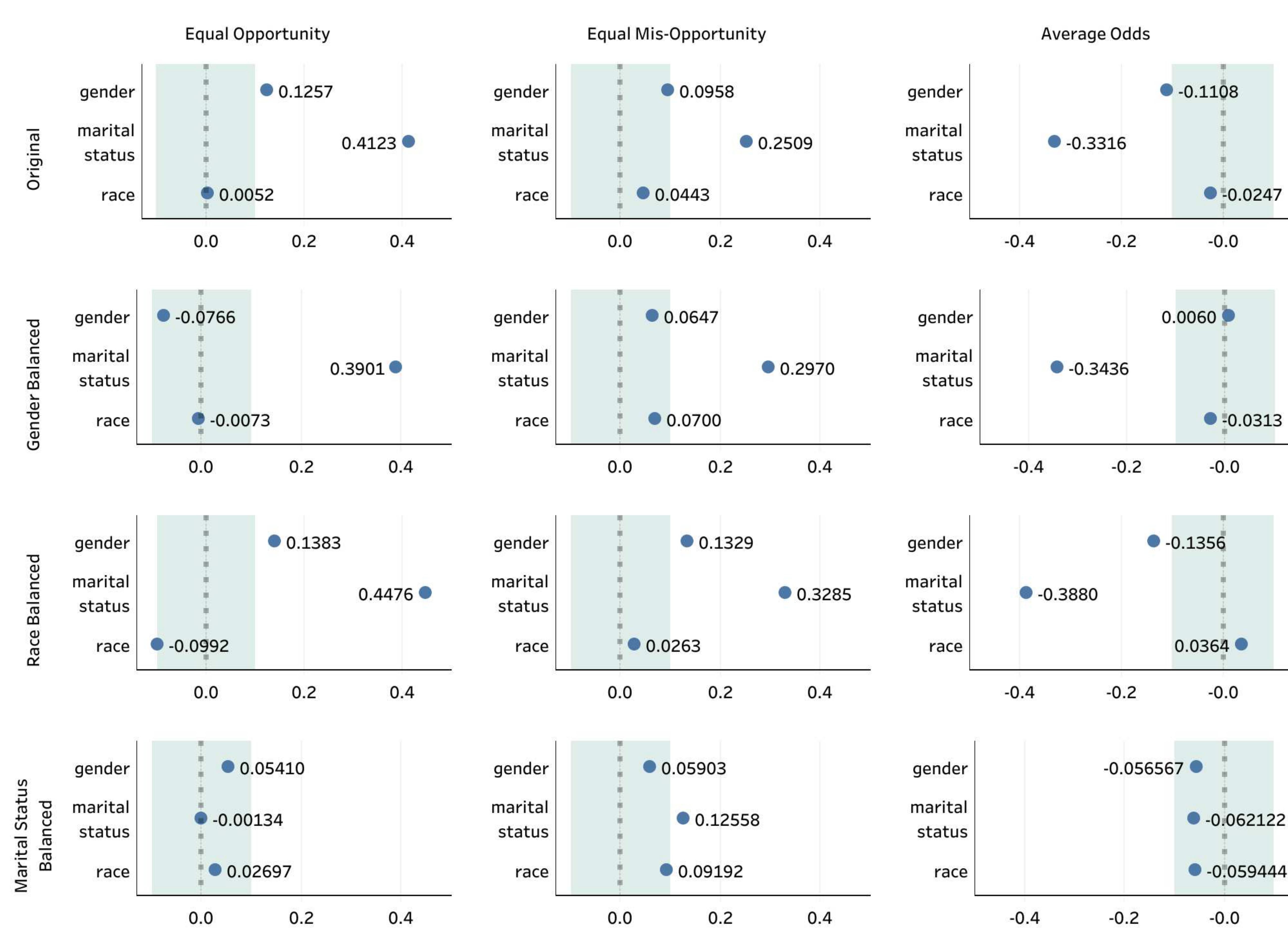}
\caption{Equal Opportunity, Equal Mis-Opportunity and Average Odds values for the predictions of the test datasets for the four scenarios of the UCI Adult dataset}
\label{fig:2}
\end{figure}

Figure \ref{fig:1} shows Statistical Parity Difference and Disparate Impact calculated on the pre-trained dataset for all four iterations. Figure \ref{fig:2} shows Equal Opportunity Difference, Equal Mis-Opportunity Difference, and Average Odds Difference calculated on the model predictions on the test dataset. As can be seen, the original dataset has a bias for all the protected attributes. Further, in subsequent iterations, when we calculate metrics by resampling for each protected attribute one by one, the values start falling in the acceptable band for that attribute. We also observe that while resampling for a protected attribute reduces the bias in that attribute, it also tends to reduce the bias for other attributes.

\subsection{German Credit Dataset}
\label{germanCreditDataset}
Each record in the German credit dataset represents an individual, classified as either a good credit risk or a bad credit risk. The dataset contains attributes like Sex, Housing, Purpose, Duration among others.

For our experiment, we considered the following configurations:
\begin{align*}
Protected \ attributes&: Sex \\
Privileged \ class&: male \ (Sex) \\
Unprivileged \ class&: female \ (Sex) \\
Favourable \ outcome&: good \ (Risk)
\end{align*}

\begin{figure}[h]
\centering
\includegraphics[width=\linewidth]{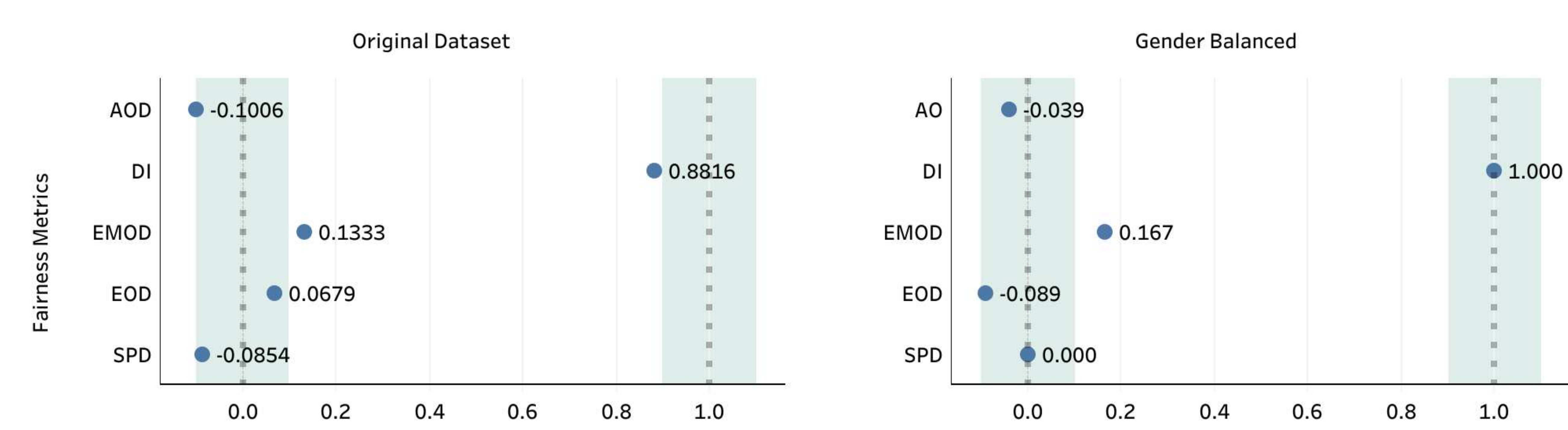}
\caption{Fairness metrics for the Original (left) and Gender-Balanced (right) German Credit Dataset}
\label{fig:3}
\end{figure}

Figure \ref{fig:3} shows the fairness metrics calculated on the original dataset and the dataset after mitigating bias in the protected attribute. The metrics indicate some gender bias in the original dataset that gets addressed in the gender-balanced dataset.

\subsection{Health Insurance Cross-Sell Prediction Dataset}
\label{healthInsuranceCrossSellPredictionDataset}
Health Insurance Cross-Sell Prediction dataset is an insurance company's dataset of customers to whom the company provided health insurance in the last one year. The dataset is used to predict the customers interested in buying the company's vehicle insurance. The dataset contains attributes like DrivingLicense, VehicleAge, VehicleDamage, AnnualPremium among others.

For our experiment, we considered the following configurations:
\begin{align*}
Protected \ attributes&: Gender \\
Privileged \ class&: Male \ (Gender) \\
Unprivileged \ class&: Female \ (Gender) \\
Favourable \ outcome&: 1 \ (Response)
\end{align*}

\begin{figure}[h]
\centering
\includegraphics[width=\linewidth]{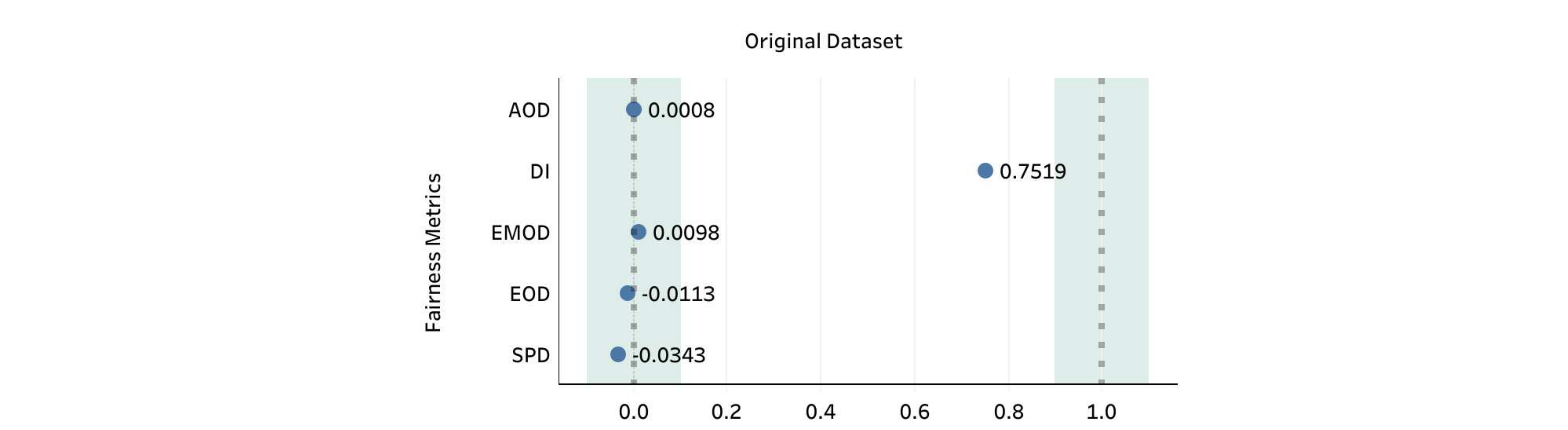}
\caption{Fairness metrics for the Health Insurance Cross-Sell Prediction Dataset}
\label{fig:4}
\end{figure}

Figure \ref{fig:4} shows the fairness metrics calculated on the original dataset. The metrics, except Disparate Impact, indicate that the dataset is free for bias for the protected attribute.

\begin{table}[h]
\small
\centering
\resizebox{\textwidth}{!}{%
\def\arraystretch{1.5}%
\begin{tabular}{c c c c c c c c}
\hline
Protected Attribute & SPD     & DI     & EOD     & EMOD   & AOD     & Bias Index & Fairness Value          \\ \hline
\multicolumn{8}{c}{UCI Adult dataset - Original dataset}                                                 \\ \hline
Race                & -0.1019 & 0.5990 & 0.0052  & 0.0443 & -0.0247 & 0.1864     & \multirow{3}{*}{0.6457} \\ 
Gender              & -0.1945 & 0.3598 & 0.1257  & 0.0958 & -0.1108 & 0.3114     &                         \\ 
Marital Status      & -0.3831 & 0.1429 & 0.4123  & 0.2509 & -0.3316 & 0.4948     &                         \\ \hline
\multicolumn{8}{c}{UCI Adult dataset - Gender balanced}                                                  \\ \hline
Race                & -0.1463 & 0.5833 & -0.0073 & 0.0700 & 0.0313  & 0.2005     & \multirow{3}{*}{0.7025} \\ 
Gender              & 0.0000  & 1.0000 & -0.0766 & 0.0647 & 0.0060  & 0.0449     &                         \\ 
Marital Status      & -0.3912 & 0.2224 & 0.3901  & 0.2970 & -0.3436 & 0.4725     &                         \\ \hline
\multicolumn{8}{c}{UCI Adult dataset - Race balanced}                                                    \\ \hline
Race                & 0.0000  & 1.0000 & -0.0992 & 0.0263 & 0.0364  & 0.0487     & \multirow{3}{*}{0.6508} \\ 
Gender              & -0.2400 & 0.4096 & 0.1383  & 0.1329 & -0.1356 & 0.3038     &                         \\ 
Marital Status      & -0.4623 & 0.1732 & 0.4476  & 0.3285 & -0.3880 & 0.5208     &                         \\ \hline
\multicolumn{8}{c}{UCI Adult dataset - Maritial Status balanced}                                         \\ \hline
Race                & -0.1144 & 0.6689 & 0.0270  & 0.0919 & -0.0594 & 0.1646     & \multirow{3}{*}{0.8909} \\ 
Gender              & 0.0380  & 0.8881 & 0.0541  & 0.0590 & -0.0566 & 0.0687     &                         \\ 
Marital Status      & 0.0000  & 1.0000 & -0.0013 & 0.1256 & -0.0621 & 0.0627     &                         \\ \hline
\multicolumn{8}{c}{German Credit dataset - Original dataset}                                             \\ \hline
Sex              & -0.0854 & 0.8816 & 0.0679  & 0.1333 & -0.1006 & 0.1037     & 0.8963                  \\ \hline
\multicolumn{8}{c}{German Credit dataset - Sex balanced}                                                 \\ \hline
Sex              & 0.0000  & 1.0000 & -0.0890 & 0.1670 & -0.0390 & 0.0864     & 0.9136                  \\ \hline
\multicolumn{8}{c}{Health Insurance dataset - Original dataset}                                          \\ \hline
Gender              & -0.0343 & 0.7519 & -0.0113 & 0.0098 & 0.0008  & 0.1122     & 0.8878                  \\ \hline
\end{tabular}%
}
\caption{Fairness Scores for the datasets, Bias Indexes and various Fairness Metrics calculated for the protected attributes for all the seven cases}
\label{table:1}
\end{table}

Fairness Scores calculated for the datasets, Bias Indexes and various Fairness Metrics calculated for the protected attributes for all the seven cases are tabulated in Table \ref{table:1}.

\section{Discussion}
\label{discussion}
Analysis of the calculated values and their graphs related to the experiment are as follows:
\begin{enumerate}
    \item The training datasets used for various AI systems are not perfect, so the AI systems are likely to be prejudiced and may not be fair in all aspects. Biases may also be there due to imperfections in the algorithms in the model. Our experiment indicates that different metrics are required to check for biases in the pre-processed training dataset and the outcomes. For the former case, we have used Statistical Parity Difference and Disparate Impact. For the latter case, we have used Equal Opportunity Difference, Equal Mis-Opportunity Difference, and Average Odds Difference.
    \item A single metric might not correctly identify bias in all cases, so using a combination of metrics is suggested. We have used two metrics for the pre-processed training dataset and three metrics for the post-training outcomes.
    \item Different metrics may be used in other AI systems, but the Fairness Score will still be relevant and can be used to compare various systems.
    \item The training dataset might have multiple attributes that might impact the fairness of the system. It is necessary to check for fairness for each protected attribute. In our experiment, we checked for fairness for race, gender, and marital status in the UCI Adult Dataset and gender in the other two datasets.
    \item Testing all the protected attributes combined is difficult, whereas checking for each attribute separately is practically implementable.
    \item Our experiment confirms bias in the protected attributes of race, gender, and marital status in the UCI Adult Dataset and gender in the German Credit Dataset. Further, no significant bias is observed for the protected attribute of gender in the Health Insurance Cross-Sell Prediction dataset. These results are validated by previous studies \citep{tramer2015discovering}.
    \item The graphical representation also shows the relative degree of bias for one attribute as compared to others. The UCI Adult Dataset has a maximum bias in the case of marital status.
    \item When bias for one attribute is removed, the impact on other protected attributes can also be checked from this graphical representation. It is observed that when bias for marital status was removed, the biases for the other two protected attributes also got reduced.
    \item As having a perfect unbiased system is impossible, a tolerance band is recommended. We have considered the band to be ±10\%. The width of this band can be decided on the use case, narrow in cases where bias is not tolerable and broader where bias is not of much significance. The Four-Fifths Rule, Uniform Guidelines on Employee Selection Procedures (1978) of the US Government considers the adverse impact on a group if its selection rate is less than four-fifths (or 80\%) of the rate of the group with the highest selection rate, thus indicating tolerance of 20\% \citep{cfr}. Accordingly, a Bias Index between 0 and 0.1 for a protected attribute indicates that the system is unbiased for that attribute. Similarly, Fairness Score between 0.9 and 1.0 indicates a fair AI system.
    \item Experiments on multiple datasets, some having bias and others balanced for bias, validate the proposed Fairness Score, the Bias Index and the SOP as these correctly assess the bias or fairness of the AI models.
\end{enumerate} 

\section{Conclusion}
\label{conclusion} 
Just as there are established procedures for security audit and certification of websites and portals by designated auditing agencies against accepted benchmarks such as OWASP Top 10, an ecosystem for fairness audit and certification for AI-based Systems also needs to evolve. Standardization of the bias determining process is essential for more acceptance of such systems. This paper proposes a Fairness Score and a Standard Operating Procedure to evaluate the fairness of such AI systems and provide a framework for Fairness Certification. Bias in an AI/ ML system can be present due to the training dataset or the algorithms used. These biases are measured using pre-defined metrics. A single metric may not be accurate for all the systems and applications, so a combination of metrics is preferred to evaluate their fairness. This paper outlines the desired features of such metrics and shortlists the suitable metrics for data-driven AI systems. As an AI/ ML system continuously learns while in operation, it is necessary to check for fairness periodically to ensure its neutrality and fair decision-making ability. Our experiments on the Census Income dataset, German Credit dataset, and Health Insurance dataset establish that the Fairness Score and the SOP proposed in the paper can be used as a framework to evaluate the fairness of AI systems and issue a Fairness Certificate.

Further studies can be carried out to extend the framework to specific use-cases and different types of AI systems. Some ethical questions arise as a result of the framework. Who decides the protected attributes of the dataset? Also, the selection of privileged and unprivileged classes can be manipulated to give a favourable result. There could be a bias or manipulation in deciding the protected attributes or the privileged/ unprivileged classes. Further, a record might belong to privileged class in one use case and unprivileged class in another, making it difficult to have strict guidelines for privileged class selection. Answers to these concerns could be considered for further research and studies.

\section{Declarations}
\label{declaration}
\textbf{Funding} The authors did not receive support from any organization for the submitted work. \\
\textbf{Availability of data and material} Not applicable. \\
\textbf{Conflicts of interest/ Competing interests} On behalf of all authors, the corresponding author states that there is no conflict of interest. \\
\textbf{Code availability} Not applicable.

\bibliographystyle{unsrtnat}

\end{document}